\documentclass[twoside]{dis09}
\usepackage[latin1]{inputenc}
\usepackage[dvips]{graphicx,epsfig,color}
\usepackage{wrapfig,rotating}
\usepackage{amssymb,amsmath,array}

\begin{document}
\title{Hadronisation Corrections for Jets in the $k_t$ Algorithm}

\author{Mrinal Dasgupta$^1$ and Yazid Delenda$^2$
%
%
\vspace{.3cm}\\
%
1- School of Physics and Astronomy - University of Manchester \\
Oxford Road, Manchester M13 9PL - United Kingdom.
%
\vspace{.1cm}\\
2- D\'{e}partement de Physique - Facult\'{e} des Sciences \\
Universit\'{e} Hadj Lakhdar, Batna 05000 - Algeria.\\
}

\maketitle \vspace{-5.0cm} \hfill MAN/HEP/2009/30 \vspace{4.55cm}

\begin{abstract}
It has recently been established that hadronisation corrections to
QCD jets vary as $1/R$, at small $R$, for jets of radius $R$. Here
we demonstrate, using jets in the $k_t$ algorithm, that the
magnitude of these $1/R$ corrections are unambiguously linked to the
magnitude of $1/Q$ corrections to commonly studied event shapes in
$e^{+}e^{-}$ annihilation.
\end{abstract}

\section{Introduction}

An understanding of QCD jets and their properties will be integral
to the success of the LHC physics program. In particular one of the
most important issues in current jet studies is the question of jet
energy scale. The shift induced in jet energy by effects such as
perturbative radiation and non-perturbative effects like
hadronisation and the underlying event would contribute to a
smearing of, for instance, mass peaks that may be a signal for new
physics. Thus in order to choose optimal jet definitions that
minimise such smearing one would need to know the dependence of the
above effects on the experimental parameters like jet radius.
Moreover, even in pure QCD studies such as the extraction of parton
distribution functions (pdfs) and the strong coupling $\alpha_s$
from jet observables like the inclusive jet cross-sections, a
knowledge of the non-perturbative contribution is important to
supplement perturbative calculations. A relatively small shift in
the transverse momentum $p_t$ of a jet, induced by hadronisation,
can result in a significant change in the inclusive jet spectrum
since we are dealing with a quantity that has a steeply falling
$p_t$ distribution.

While it is traditional to study the hadronisation contribution via
Monte Carlo models such as those in {\tt HERWIG} and {\tt PYTHIA},
it turns out that in cases like the jet energy there is additionally
valuable analytical insight available \cite{Dasgupta:2007wa}.
Analytical models based on renormalons \cite{Beneke:1998ui} have in
the past been met with great success in the description of LEP and
HERA event-shape variables \cite{Dasgupta:2003iq}, but have not
really been utilised outside that context. In
Ref.~\cite{Dasgupta:2007wa} one such model (due to Dokshitzer and
Webber \cite{Dokshitzer:1995zt}) was used to estimate hadronisation
corrections to jet transverse momentum $p_t$. The result found there
was striking: hadronisation effects have a singular $1/R$ dependence
on the jet radius $R$, at small $R$. This is in complete contrast to
the contribution from the underlying event which varies as $R^2$.
The knowledge of the $R$ dependence of non-perturbative effects in
conjunction with the $\ln R$ behaviour involved in perturbative
estimates can then be used to arrive at conclusions about the
optimal values of $R$ to be used in diverse studies involving jets,
as exemplified in Ref.~\cite{Dasgupta:2007wa}.

While the computations of Ref.~\cite{Dasgupta:2007wa} indicate the
dependence of hadronisation corrections and the underlying event on
$R$, there remains the question of the overall magnitude of these
effects. While for the underlying event one is reliant solely on
Monte Carlo event generators to obtain the overall magnitude, for
the hadronisation correction a tentative link was made in
Ref.~\cite{Dasgupta:2007wa} between the magnitude of the $1/R$
correction and that of $1/Q$ corrections to LEP and HERA event
shapes such as the thrust distribution (see \cite{Dasgupta:2003iq}
for a review). In order to definitely link the magnitude of jet
hadronisation to that of event-shape power corrections one needs to
carry out a calculation at the two-loop level rather than the simple
one-loop estimate reported in \cite{Dasgupta:2007wa}. The
calculation for jets defined in the $k_t$ algorithm \cite{ESkt,inclukt} 
is reported here while the corresponding result for jets in the anti-$k_t$ algorithm
 is already known \cite{Cacciari:2008gp}. Work on the other jet
algorithms is currently in progress.

\section{The single-gluon result}

In the Dokshitzer-Webber model non-perturbative hadronisation
corrections are associated to the emission of a soft gluon with
transverse momentum $k_t \sim \Lambda_{\mathrm{QCD}}$. For the jet
$p_t$ case we work out the change in transverse momentum $\delta
p_t$ induced by the emission of such a gluon and combined with the
gluon emission probability (as given by \emph{perturbative} QCD)
this yields the average shift in $p_t$ induced by hadronisation:
\begin{equation}
\label{eq:oneor_delenda} \langle \delta p_t \rangle^h \sim
\frac{C_j}{ 2 \pi} \int \frac{dk_t}{k_t} d\eta \frac{d\phi}{2\pi}\,
\delta p_t(k) \, \alpha_s(k_t),
\end{equation}
where $k_t$, $\eta$ and $\phi$ respectively denote the transverse
momentum, rapidity and azimuth of the emitted gluon with respect to
the emitting hard parton (jet) and $C_j$ is the colour factor
associated to emission from the given hard parton. The reason we are
able to single out a given hard parton initiating a high-$p_t$ jet
in any hard process is essentially since the leading $1/R$ result
stems from emission collinear to the triggered jet. It is thus
possible to talk in terms of the single-jet limit ignoring the rest
of the details of the hard process.

The only non-perturbative ingredient that is involved above is the
value of $\alpha_s(k_t)$ at scales around or below $\Lambda_{
\mathrm{QCD}}$. If one makes the assumption of a universal
infrared-finite coupling, which replaces the perturbative coupling
that has an unphysical divergence at $\Lambda_{\mathrm{QCD}}$, then
one arrives at a prediction for the hadronisation correction
$\langle \delta p_t \rangle^h$. Using the fact
\cite{Dasgupta:2007wa} that the $\delta p_t(k)$ is essentially the
energy of the gluon emitted outside the jet we perform the integral
over rapidity in Eq.~\eqref{eq:oneor_delenda} to obtain the leading
$1/R$ behaviour. The result is of the form $c \, {\mathcal{A}}/R $,
where $c$ is a number obtained from the rapidity integral and
${\mathcal{A}}$ is the moment of the coupling $\alpha_s(k_t)$ over
the infrared region (we refer the reader to
Refs.~\cite{Dokshitzer:1997iz,Dokshitzer:1998pt} for the precise
details). Since the same coupling moment enters the predictions for
event-shape variables we can take its value from data on event
shapes and hence obtain a numerical prediction for the leading $1/R$
hadronisation correction to jet $p_t$. This was the method adopted
in Ref.~\cite{Dasgupta:2007wa}.

Here we point out a limitation of the above approach
\cite{Nason:1995hd} which is that while we have written down and
used a running coupling $\alpha_s (k_t)$, this quantity only emerges
when one considers not just the emission of a single gluon but in
fact gluon decay as well. To be precise an inclusive integration
over gluon decay products is responsible for building up the
quantity $\alpha_s(k_t)$. Unfortunately, as is known for event-shape
variables, our observable is sensitive to the precise details of
gluon branching and hence one is not free to carry out such an
inclusive integration. One must therefore return to the details of
the gluon branching and identify the correction to the above
inclusive approximation. The analysis at this level has already been
carried out for event-shape variables
\cite{Dokshitzer:1997iz,Dokshitzer:1998pt,Dasgupta:1999mb,Dasgupta:1998xt}
and below we report on it for the jet $p_t$ case.

\section{Non-perturbative effects and gluon decay}

\begin{wrapfigure}{r}{0.5\columnwidth}
\centerline{\includegraphics[width=0.45\columnwidth]{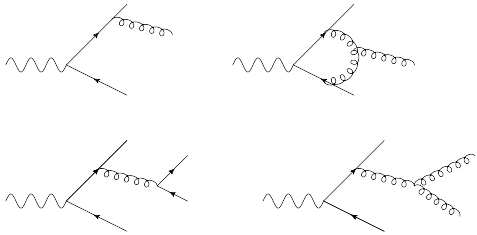}}
\caption{Gluon decay and one-loop corrections to single-gluon
emission.}\label{fig:nlo_delenda}
\end{wrapfigure}

Now we consider the situation where the emitted gluon with $k_t \sim
\Lambda_{\mathrm{QCD}}$ is allowed to decay and at the same accuracy
account for virtual corrections to single gluon emission, as
depicted in Fig.~\ref{fig:nlo_delenda}.

At this two-loop level the change in $p_t$ (for a quark jet) can be
expressed as:
\begin{equation}
\langle \delta p_t \rangle^h = \frac{C_F}{\pi} \int \frac{d^2
k_t}{\pi k_t^2} \frac{d\alpha}{\alpha} \left \{\alpha_s(0)+4 \pi
\chi (k_t^2) \right \} \delta p_t (k) +4 C_F \int \left (
\frac{\alpha_s}{4 \pi} \right)^2 d \Gamma_2 \frac{M^2}{2!} \delta
p_t (k_1,k_2),\nonumber
\end{equation}
where $\alpha$ is a Sudakov variable, $\alpha_s(0)$ is an
ill-defined quantity which will cancel away subsequently, $\chi$
represents the virtual correction to gluon emission, $d\Gamma_2$ is
the gluon decay phase-space and $M^2$ is the decay matrix element
\cite{Dokshitzer:1997iz,Dokshitzer:1998pt}. We also denote by
$\delta p_t(k_1,k_2)$ the change in $p_t$ due to correlated
two-parton emission while $\delta p_t(k)$ is the corresponding
single-gluon quantity. To correctly account for gluon branching one
thus has to perform the above calculation, the details of which are
reported in Ref.~\cite{Dasgupta:2009tm}.

The analogous two-loop analysis for event-shape variables
\cite{Dokshitzer:1997iz, Dokshitzer:1998pt, Dasgupta:1999mb,
Dasgupta:1998xt} revealed an initially surprising result -- the
two-loop correction simply provided a \emph{universal} rescaling
factor to the one-gluon result, which became known as the
\emph{Milan factor}. Its value for $n_f=3$ (which is the number of
flavours excited in the relevant soft region) was found to be
$\mathcal{M} = 1.49$. Thus the ratio of corrections to two event
shapes $v_1$ and $v_2$ was merely the ratio of the one-loop
coefficients computed previously \cite{Dokshitzer:1995zt}:
\begin{equation*}
\frac{\delta v_2^{\mathrm{NP}}}{\delta v_1^{\mathrm{NP}}}=
\frac{\delta v_2^{\mathrm{NP},1}}{ \delta v_1^{\mathrm{NP},1}},
\end{equation*}
where $\delta v_1^{\mathrm{NP},1}$ denotes the non-perturbative
single-gluon correction for $v_1$ computed as discussed in the
preceding section and likewise for $v_2$. This remarkable result was
understood to arise as a consequence of the fact that all the
variables considered could be expressed as linear sums over the
transverse momenta $k_{ti}$ of emissions, $v =\sum_i k_{ti} c_i$,
where the $c_i$ are rapidity-dependent coefficients
\cite{Dokshitzer:1998pt}.

\begin{wrapfigure}{r}{0.5\columnwidth}
\centerline{\includegraphics[width=0.45\columnwidth]{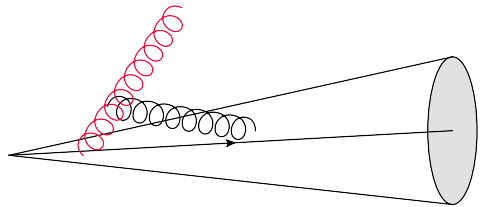}}
\caption{The role of clustering in determining whether a soft gluon
ends up within or outside a hard jet.}\label{fig:cluster_delenda}
\end{wrapfigure}

In the case of jet $p_t$ this linear dependence is ruined by the
non-trivial action of the jet algorithm in all cases except the case
of jets defined in the anti-$k_t$ algorithm \cite{Cacciari:2008gp}.
The contribution to the jet $\delta p_t$ of a given emission is
found to be of the form $k_t e^\eta\, \Xi_{\mathrm{out}}$, where
$\eta$ denotes the rapidity with respect to the emitting hard jet
and $\Xi_{\mathrm{out}}$ denotes the condition that the emission
ends up outside the jet \emph{after the application of the jet
algorithm}. It should be immediately clear from this that in most
current sensible jet algorithms (both of sequential recombination
and cone type) the condition $\Xi_{\mathrm{out}}$ is non-trivial and
introduces non-linearity in $k_t$. For instance in the $k_t$
algorithm we can consider the situation in
Fig.~\ref{fig:cluster_delenda}, where although one may have a soft
parton separated by more than a certain distance $R$ in rapidity and
azimuth (denoted by the red gluon line) from a given hard parton, it
may be clustered to another soft parton (denoted by the black gluon
line) and hence swept into the final jet. This clustering depends on
the $k_t$ of a soft parton relative to the other partons and hence
the condition $\Xi_{\mathrm{out}}$ derived in \cite{Dasgupta:2009tm}
contains dependence on the $k_t$ of the soft partons, spoiling the
simple linear dependence needed for universality.

An exception to the above situation is to be found in the anti-$k_t$
algorithm for which the condition $\Xi_{\mathrm{out}} = \Theta
(\eta^2+\phi^2-R^2)$ ensures that a given parton is outside the jet
if its angular ($\eta,\phi$) separation from the hard jet is more
than $R$. The linear dependence on $k_t$ is maintained and the Milan
factor $\mathcal{M}=1.49$ is computed as for event shapes.

For the $k_t$ algorithm we have carried out an equivalent
calculation for the leading $1/R$ hadronisation correction (at small
$R$) with the more complicated $\Xi_{\mathrm{out}}$ function
involved there and we found the result $\mathcal{M}_{k_t}=1.01 \,
(n_f=3)$. Thus while at the level of the one-gluon studies of
Ref.~\cite{Dasgupta:2007wa} the $k_t$ and anti-$k_t$ algorithms
received identical hadronisation corrections, a detailed analysis at
the two-loop level breaks this equality. One finds that the ratio of
hadronisation corrections is then:
\begin{equation}
\frac{{\langle \delta p_t \rangle}^h_{k_t}}{{\langle\delta p_t
\rangle}^h_{\mathrm{anti-}k_t}} = \frac{1.01}{1.49} \sim
0.7\,.\nonumber
\end{equation}

Thus one expects somewhat smaller hadronisation corrections for the
$k_t$ algorithm as compared to those for the anti-$k_t$ algorithm
which is also borne out by the Monte Carlo studies with {\tt HERWIG}
and {\tt PYTHIA} reported in Ref.~\cite{Dasgupta:2007wa}. We remind
the reader that these conclusions apply only to the $1/R$
hadronisation corrections that would be dominant at small $R$ and we
neglect finite $R$ corrections which need to be considered alongside
the underlying event contribution which also has a regular $R$
dependence $\sim R^2$.

\section{Conclusions}

We have reported on a study of hadronisation corrections to jet
$p_t$ or energy scale based on two-loop extensions of the one-gluon
estimates reported in Ref.~\cite{Dasgupta:2007wa},  with jets
defined in the $k_t$ algorithm. Studies for other jet algorithms
(SISCone \cite{Salam:2007xv} and Cambridge/Aachen
\cite{Dokshitzer:1997in-Wobisch:1998wt}) are in progress.


\begin{footnotesize}



%

\end{footnotesize}


\end{document}